# Enhancing Higher Education with Generative AI: A Multimodal Approach for Personalised Learning

Johnny Chan[0000-0002-3535-4533] and Yuming Li[0000-0003-2219-9033]

The University of Auckland, Auckland, New Zealand
jh.chan@auckland.ac.nz, yuming.li@auckland.ac.nz

**Abstract.** This research explores the opportunities of Generative AI (GenAI) in the realm of higher education through the design and development of a multimodal chatbot for an undergraduate course. Leveraging the ChatGPT API for nuanced text-based interactions and Google Bard for advanced image analysis and diagram-to-code conversions, we showcase the potential of GenAI in addressing a broad spectrum of educational queries. Additionally, the chatbot presents a file-based analyser designed for educators, offering deep insights into student feedback via sentiment and emotion analysis, and summarising course evaluations with key metrics. These combinations highlight the crucial role of multimodal conversational AI in enhancing teaching and learning processes, promising significant advancements in educational adaptability, engagement, and feedback analysis. By demonstrating a practical web application, this research underlines the imperative for integrating GenAI technologies to foster more dynamic and responsive educational environments, ultimately contributing to improved educational outcomes and pedagogical strategies.

## 1    Background

With the rapid advancement of artificial intelligence (AI) technologies, particularly the emergence of Generative AI (GenAI), both academia and industry are witnessing a new wave of innovation. GenAI has demonstrated immense potential and applicability across various sectors, including education, healthcare, and entertainment [1]. In the domain of education, especially higher education, there is an increasing demand for personalised learning experience. Students seek learning methods that are more customised, interactive, and flexible, which traditional educational tools and methods struggle to provide [2]. To address the challenge, this paper introduces a multimodal conversational chatbot based on GenAI technology, designed to significantly enhance the learning experience in higher education through intelligent, personalised dialogue interactions.

In the realm of educational technology, despite the advent of GenAI tools including ChatGPT's Scholar Pro and Academic Assistant plugins [3], these often cater to a broad range of topics. Specific to different disciplines, some higher education institutions integrate conversational bots into their learning management systems to provide text-based interactions for information retrieval and course assistance. However, this unimodal method of interaction overlooks an important aspect in educational settings: the input and processing of non-textual information, especially visual information such as images. In an actual learning environment, students often need to work with images, like some diagrams from the lecture slides or some scientific illustrations from the course materials, which are the type of information that traditional text-based AI assistance tools struggle to effectively handle. Moreover, existing studies and products are nearly void in the specific direction of converting diagram to code. This conversion is not only crucial for disciplines like computer science, information systems, data science, and engineering but also holds immense value for any field involving graphical representation and programming. Currently, although the conversion from code to diagram is relatively mature [4] as it mainly involves the compilation process, the conversion from diagram to code faces significant challenges. Firstly, it requires highly precise image recognition technologies to accurately understand the symbols, structures, and semantic information presented in the diagram. Secondly, the diversity in style, format, and detail of diagram produced by different software tools would require the conversion tool be capable of processing diagrams from various sources uniformly into standard code, which is a technically difficult task.

Furthermore, among the current educational assistance technologies, the absence of file input functionality represents another limitation. This feature is crucial for achieving deeper personalised learning and enhancing teaching efficiency. By supporting file input, students and teachers can directly upload documents relevant to the course, such as course-related materials and course evaluation reports, thereby obtaining customised feedback, analysis reports, or other forms of educational assistance. This capability can not only help students better understand the learning materials and identify weaknesses in their learning but also enable teachers to efficiently manage and respond to students' learning needs, providing timely course optimisation and personalised instructional support.

Based on the aforementioned background, this paper proposes a multimodal conversational chatbot based on GenAI, aimed at enhancing the personalised learning experience in higher education. The main contribution of this conversational chatbot lies in its support for multimodal inputs, including text, images, and files, enabling it to process and respond to more complex and diversified education-related queries. Specifically, this system is not only capable of understanding and responding to text-based student questions but can also process images uploaded by students (such as classroom PowerPoint screenshots, scientific diagrams, etc.) and various educational documents (e.g., student course assessment reports), thereby providing accurate and relevant feedback and answers. Crucially, our system offers a solution to the diagram-to-code conversion challenge, a problem not yet widely addressed in the educational technology field. This capability is particularly applicable to STEM courses, helping students and teachers to more effectively understand and utilise diagrams and their corresponding code implementations, thus enhancing teaching and learning efficiency. Furthermore, by supporting file input, our system further expands its application scope, offering personalised guidance on students' learning materials, as well as automating the analysis of students' homework and assessment reports for teachers, thereby optimising the teaching process and improving the quality of instruction.

## 2      Literature Review

In recent years, the domain of educational technology, particularly in enhancing learning experiences through the use of AI, has witnessed significant achievements in applications based on text. These accomplishments are primarily manifested in leveraging natural language processing (NLP) technologies to understand and generate textual content, thereby providing personalised learning experiences, automated assessments, and real-time feedback [5], [6], [7]. These studies span from rule-based systems to deep learning based chatbots, capable of offering user-oriented learning resources, answering academic questions, and facilitating understanding and retention based on students' inputs. Although text-centric applications have maturely developed, the exploration and application of multimodal conversational AI in the educational sector are still in their early stages [2]. Multimodal conversational AI refers to artificial intelligence systems capable of processing and understanding various forms of input, such as text, images, and voices. The integration of this technology can greatly enrich the interactivity and adaptability of educational applications, offering students a more comprehensive and engaging learning experience.

In the research on constructing multimodal AI chatbots, Lee [8] expanded the discussion on integrating visual information into chatbots. The research distinguishes between image-based dialogue and enhanced image dialogue, emphasising the importance of incorporating images as supplementary components to enrich dialogue content. It also points out that the ability of chatbots to understand and engage with a broader range of inputs is evolving, not limited to text but also including images and potentially other forms of data.

In summary, the transformation towards intelligent education driven by advancements in GenAI technology is an inevitable trend [9]. Although the exploration of integrating multimodal conversational AI into education is still in its infancy, it has already shown tremendous potential. With the development of technology and further research, future educational chatbots are expected to evolve beyond mere text-based communicator to intelligent teaching assistant capable to comprehensively understand and respond to a variety of modal inputs. This will not only enhance the quality and efficiency of the learning experience but also provide students with richer and more diverse learning pathways.

## 3      Methodology

This paper aims to enhance personalised learning experiences in higher education by developing a multimodal chatbot powered by GenAI technology. The core of this research lies in its integration of multimodal input functionalities for text, images, and files, the implementation of a diagram-to-code feature with a file analysis component.

### 3.1      Text-based Chatbot Design

In the text-based Q&A module, we have adapted the ChatGPT API to specific course through fine-tuning. Fine-tuning, a principle based on pre-trained language models like ChatGPT, involves further training the model on a specific task to adjust its parameters, making it better suited to the requirements of that task. The data format typically includes pairs of input text and the desired output text. In this module, data drawn from past exams and classroom quizzes, which have been anonymised, are used to train the model to generate answers tailored to specific educational content. To adhere to the requirements of double-blind reviewing, specific data details are not disclosed extensively. The format is as follows:

**{"prompt": "<prompt text>", "completion": "<ideal generated text>"}**

Where the "prompt" field contains the input text, such as a question or a prompt, and the "completion" field contains the ideal text that the model should generate, namely the answer to the question or an expansion of the prompt. This format helps the model learn how to generate the desired output based on specific inputs, thereby improving its performance on the particular task.

### 3.2      Image-based Chatbot Design

In the image-based Q&A module, we utilise the API of Google Bard to interpret user-uploaded images. Compared to ChatGPT, Google Bard possesses strong capabilities in analysing scientific and engineering images, especially in recognising elements and structures within diagrams [10]. This allows an effective conversion of image information into relevant code or textual explanations. Hence, we chose it as the ideal option for image-based question answering and diagram-to-code conversion in this study.

### 3.3      File based Analyser Design

In the file-based analyser module, we support the upload of PDF files, like course reading materials or course evaluation reports, aimed for teaching staff. This analyser will provide functionalities and data analysis including sentiment distribution, sentiment score distribution, comment length distribution, comment word count distribution, emotion distribution (based on Plutchik's wheel of emotions, which identifies eight basic emotions), top 20 keywords, and comment summary. When analysing course evaluations, by employing NLP technologies and machine learning models, key indicators such as emotional distribution, sentiment scores, lengths of comments, and word count distributions can be extracted from student feedback. These analyses help educators gain a deep understanding of student satisfaction, emotional fluctuations, and key concerns. For instance, sentiment scores reveal students' overall feelings towards a course, while the application of Plutchik's emotion wheel further refines this emotion analysis by identifying the basic emotions present in the feedback. Moreover, keyword extraction and comment summaries allow teaching staff to quickly grasp the main content of the feedback, providing data support for enhancing teaching quality and course design.

## 4     Proof-of-concept

Our proof-of-concept is constructed using Gradio [11], an open-source library designed for rapidly creating web application interfaces for machine learning models. It supports a variety of input and output formats, making it suitable for multimodal applications.

In the text-based chatbot module, users can engage in instant dialogue about course-related content. They input their questions into the dialogue box and click 'submit' to send, as shown in Fig. 1:

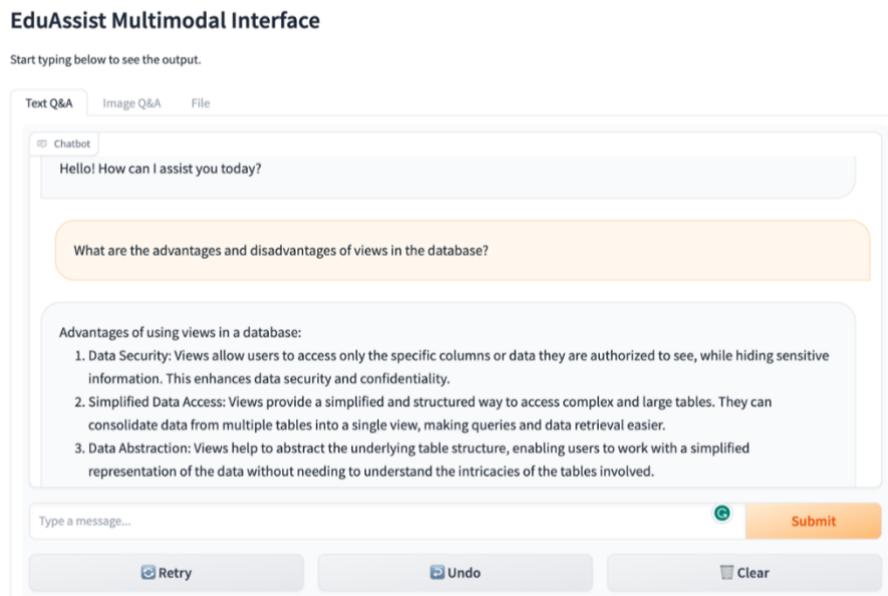

**Fig. 1.** The demonstration of test-based chatbot module

In the image-based chatbot module, we support queries related to uploaded images. For instance, as shown in Fig. 2, our proof-of-concept provides a descriptive analysis for an uploaded pie chart.

**Fig. 2.** The demonstration of image-based chatbot module

Notably, our image-based chatbot supports a diagram-to-code feature. Specifically, we showcase the conversion of a typical entity-relationship diagram (ERD) into structure query language (SQL) code from a undergraduate database course as an example.

**Fig. 3.** The demonstration of image-based chatbot module: diagram-to-code

From Fig. 3, it is evident that the accuracy of our image-based chatbot module in converting ERDs to code is fundamentally accurate.

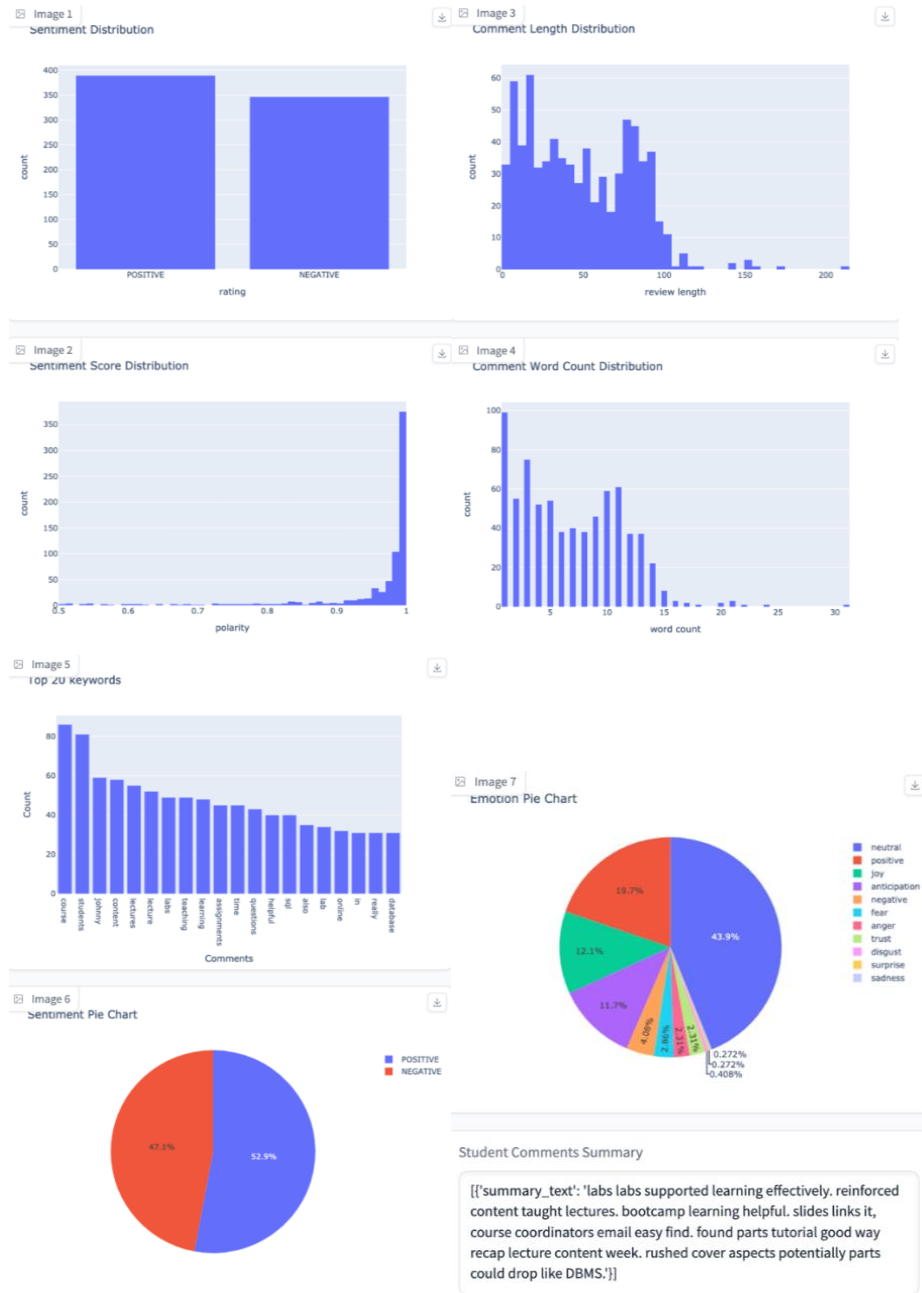

**Fig. 4.** The demonstration of file-based analyser module

In the file-based analyser module, we input the summative evaluation of a course from a higher education institution (without disclosing specific contents), and the results reveal the distribution of student sentiments, the scale of student comments, as well as keywords and summaries, as shown in Fig. 4.

Such an analysis tool is extremely valuable to educators. This analyser not only quantifies the tendencies of student comments but also enhances the efficiency of teaching iterations. For example, in large courses where the number of students can reach hundreds, manually analysing course evaluation feedback becomes impractical. Using a file-based analyser to automatically process these evaluations can significantly save time, quickly and accurately reveal student sentiment distribution, comment scale distribution, keywords, and summaries of course feedback, thus providing a strong basis for teaching improvement.

## 5     Conclusion

This paper explores the transformative potential of GenAI technologies in enhancing personalised learning experiences within higher education. By developing a multimodal chatbot that integrates text, images, and file input, this research addresses the need for more interactive and adaptable educational tools. The chatbot's capabilities, including text-based Q&A with fine-tuning via the ChatGPT API, image-based Q&A powered by Google Bard with diagram-to-code feature, exemplify the application of GenAI in catering to diverse educational inquiries and tasks. Furthermore, the paper also introduces a comprehensive file-based analyser designed for educators, offering insightful metrics such as sentiment and emotion distributions, comment length, and keyword summaries from course evaluations. This tool demonstrates significant benefits in understanding student feedback at scale, enhancing the efficiency of teaching iterations, and informing course improvements. The findings of this research underscore the evolving role of multimodal conversational AI in education, promising not only to enrich the learning experience but also to provide educators with powerful tools for assessment and course development. This research sets a foundational step towards a future where educational chatbots and analysers, powered by GenAI, offer nuanced, comprehensive, and personalised support to both students and educators alike.